\documentclass[aps,pre,reprint,twocolumn,superscriptaddress,showpacs]{revtex4-2}
\usepackage{amssymb,amsmath}
\usepackage[table]{xcolor}
\usepackage{graphicx}
\usepackage{mathtools}
\usepackage[export]{adjustbox}
\usepackage{overpic}
\usepackage{nicefrac}
\usepackage{multirow}
\usepackage{lipsum} 
\usepackage[normalem]{ulem}
\usepackage{pbox}
\newcolumntype{C}[1]{>{\centering\let\newline\\\arraybackslash\hspace{0pt}}m{#1}}
\usepackage{verbatim}
\usepackage{enumitem}
\usepackage{caption}
\usepackage{subcaption}
\usepackage{float}
\usepackage{framed,color}
\usepackage{verbatim}

\usepackage{bbm}

\newcommand{\id}{{\openone}}

\allowdisplaybreaks

\begin{document}

\title{Local density of states distribution and multifractal
eigenvectors\\ of weighted random networks via the cavity approach}
\author{Joseph W. Baron}
\email{jwb96@bath.ac.uk}
\affiliation{Department of Mathematical Sciences, University of Bath, Bath, BA2 7AY, UK}
\author{Tim Rogers}
\email{ma3tcr@bath.ac.uk}
\affiliation{Department of Mathematical Sciences, University of Bath, Bath, BA2 7AY, UK}

\begin{abstract}
We study the local density of states (LDoS) distribution of a general class of weighted Erd\H{o}s-R\'enyi graphs. Using the cavity method, we obtain a good approximation to the full LDoS distribution and compact expressions for its power-law tails, which we show to have exponent $3$ in the extended phase. We deduce that the eigenvectors in the continuous part of the spectrum are extended but (weakly) multifractal, and we extract expressions for the associated fractal dimensions and the singularity spectrum. We also demonstrate that the inverse participation ratio in this multifractal phase exhibits an unusual logarithmic scaling with system size, which is neither fully-extended nor localised by the usual definitions. Finally, we verify that some symmetry properties (derived from the non-linear sigma model), which have been shown to hold for many systems exhibiting multifractality, also hold in our case, both for the LDoS distribution and the singularity spectrum.
\end{abstract}

\maketitle

\section{Introduction}

Anderson localisation remains one of the foremost examples of a disorder-induced phenomenon \cite{anderson1958absence, evers2008anderson}. The classic Anderson tight-binding model exhibits a transition from a `conducting' phase to an `insulating' phase at sufficiently strong on-site disorder. That is, the quantum states of non-interacting spinless fermions in this model go from being distributed statistically evenly across sites to being exponentially localised around single sites. At the critical point between these phases, multifractality and non-ergodic behaviour can be observed \cite{mirlin2000multifractality, mirlin1994distribution, mirlin2006exact}. Recently, evidence has emerged that these phenomena might also occur deep within the conducting phase. This means that extended/metallic states can have the curious property of occupying a vanishing fraction of the state-space volume in the thermodynamic limit while the absolute number of occupied sites diverges, and they can be characterised by a spectrum of fractal dimensions \cite{evers2008anderson, fyodorov2010multifractality, monthus2011anderson, monthus2017statistical}. 

Non-ergodicity of states is also a characteristic of many-body localisation \cite{alet2018many, luitz2017ergodic}, in which interacting quantum systems exhibit an out-of-equilibrium transition from an ergodic metal at low disorder to a non-ergodic insulator at strong disorder, interpreted as localisation in Fock space \cite{tarzia2020many, roy2020fock}. With the advent of low temperature experiments in recent decades, the existence of many-body localised states has now been well established theoretically in 1D random spin chains \cite{imbrie2016diagonalization, imbrie2016many}, and experimental signatures have been observed in ultracold atomic gases \cite{schreiber2015observation, bordia2016coupling, smith2016many, choi2016exploring}. An important correspondence between the many-body Fock space of the interacting quantum system and the single-particle Anderson model was postulated in the seminal work of \cite{altshuler1997quasiparticle}. Non-ergodic extended states (also known as `bad metal' behaviour) in the many-body metallic regime have been hinted at, with characteristic subdiffusive behaviour having been found in the extended phase of many-body systems \cite{luitz2017ergodic, agarwal2017rare, biroli2017delocalized}. Insight into the origin of multifractality in Anderson-type models, or random matrix models more generally, is therefore expected to reveal something of the possible mechanism for non-ergodicity in the many-body analogue. 

Aside from the prominent applications of Anderson and many-body localisation, the nature of the eigenvector statistics has become a central topic of interest in random matrix theory writ large \cite{erdHos2020random}. This has become especially true with the invention of more exotic random matrix ensembles in various modern applications, which break rotational symmetry \cite{baron2025classes}, and thus may deviate from the usual universal eigenvalue and eigenvector statistics \cite{fyodorov2013universal, tao2012random, efetov1983supersymmetry}. Accordingly, the possibility of non-ergodicity and multifractality has been investigated in many variations of the single-particle Anderson model on the Bethe lattice and random regular graphs \cite{biroli2018delocalization, monthus2011anderson, kravtsov2018nonergodic, tikhonov2016fractality, sonner2017multifractality, de2014anderson}. It has also been investigated in critical random power-law banded matrices \cite{mirlin1996transition}, generalised $\beta$-ensembles \cite{das2024robust}, the adjacency and Laplacian matrices of complex networks \cite{da2025spectral, tapias2023multifractality, silva2022analytic}, and in the generalised Rosenzweig-Porter model (with both Gaussian and heavy-tailed off-diagonal elements) \cite{biroli2021levy, venturelli2023replica, safonova2025spectral, tarzia2020many, khaymovich2020fragile, khaymovich2021dynamical}. 

In this work, we examine analytically the origin of multifractality and non-ergodicity in weighted Erd\H{o}s-R\'enyi graphs, which were shown numerically in Ref. \cite{cugliandolo2024multifractal} to exhibit multifractal behaviour. Starting from the cavity equations \cite{rogers2008cavity, susca2021cavity}, we are able to derive simple approximation schemes for the distribution of the local density of states (LDoS), and then extract information about the localisation and multifractality of the eigenvectors. We demonstrate that, as long as the squares of the edge weights have a distribution with non-zero probability mass at zero, the LDoS distribution has the same power-law tail behaviour with universal exponent $3$. This remains true in the presence of moderate on-site/diagonal disorder. 

Using this result, we demonstrate that the eigenvectors in the continuous part of the eigenvalue spectrum are extended but (weakly) non-ergodic and multifractal, with an IPR of $I_2 \sim N^{-1}\ln N$. This constitutes a middle-ground between the usual extended ($I_2 \sim N^{-1}$) and localised ($I_2 \sim N^{0}$) behaviours. We also extract the so-called singularity spectrum and fractal dimensions, seeing there the signature of multifractality as well. These results apply generally to variety of edge-weight distributions, and we verify this numerically.

The rest of the text is arranged as follows. In Section \ref{section:model}, we introduce the class of random matrix ensembles at the centre of our analysis, as well as important quantities like the local density of states, singularity spectrum and fractal dimensions. In Section \ref{section:ldos}, we describe our general strategy for quantifying the behaviour of the LDoS via the cavity approach. In Section \ref{section:ldosdist}, we perform a rare-event analysis of our expression for the LDoS distribution to understand its power-law tail. Using this power-law behaviour, we deduce the singularity spectrum and fractal dimensions in Sections \ref{section:singularityspectrum} and \ref{section:nonergodicity} respectively, and we use this information to understand in what sense the eigenvectors are non-ergodic. We also discuss how the behaviour is modified at the approach to the edge of the continuous part of the eigenvalue spectrum in Section \ref{section:edge}, and finally in Section \ref{section:summary} we conclude.

\section{Random matrix ensembles and quantities of interest}\label{section:model}
We consider a general ensemble of weighted Erd\H{o}s-R\'enyi (ER) graphs of size $N$ whose weighted adjacency matrices we denote $\underline{\underline{J}}$. The elements of this matrix can be decomposed as a direct product $J_{ij}= W_{ij} A_{ij}$, where $W_{ij} = W_{ji}$ are the weights and $A_{ij} \in \{0,1\}$ are the elements of the unweighted ER graph. As usual, $A_{ij} = A_{ji}$ is 1 with probability $p/N$ and 0 otherwise independently of the existence of any of the other edges. To rule out any trivial effects coming from disconnected components of the graph, we examine only the giant component of the network. Weights are drawn according to $W_{ij}=S_{ij}\sqrt{u_{ij}}$ where $S_{ij}$ is a random sign $\pm1$ and the $u_{ij}$ are positive and IID. For numerical results we will use uniform, half-Gaussian and half-exponential distributions each with average $\langle u\rangle=1/p$ so that the limit $p \to \infty$ yields the usual Wigner semicircle law. Additionally, for technical reasons that we elucidate in Section \ref{section:breakdown}, we are restricted to considering distributions for $u$ with a positive probability density at zero. 

The statistics of the weighted network manifest in our calculations via the non-vanishing higher-order statistics of the matrix elements. Specifically, we define 
\begin{align}
\langle J_{ij}^2 \rangle = \frac{1}{N}, \,\, \langle J_{ij}^4\rangle = \frac{\beta_4}{pN}, \,\, \cdots\,\, \langle J_{ij}^{2k}\rangle = \frac{\beta_{2k}}{p^{k-1}N}. \label{matstats}
\end{align}
Here $p>1$ controls the speed of decay of higher order moments, while the coefficients $\beta_{2k}$ are all assumed to be order one and independent of $N$ and $p$. We thus see that the random matrices $\underline{\underline{J}}$ in this work have fourth-and-higher moments of the same order in $N$ as the variance. That is, we should not expect the matrix $\underline{\underline{J}}$ to have similar properties to a Gaussian random matrix for $p \sim N^0$. Indeed, the eigenvalue statistics of random matrices of this kind were previously considered in Ref. \cite{baron2025classes}, where the deviation from Wigner's semi-circle was characterised, as were the non-trivial long-range eigenvalue correlations and non-zero spectral compressibility. Ultimately, it will be the non-negligible higher-order statistics of $J_{ij}$ that will give rise to a non-trivial LDoS distribution here. Later, we also consider the possibility of the diagonal elements $J_{ii}$ being drawn from a uniform distribution, but for now we imagine that $J_{ii}=0$. 

Now that we have discussed the ensembles of interest, let us detail the targets of our investigation, which are all quantities derived from the eigenvectors/eigenvalues of the matrix $\underline{\underline{J}}$. We denote the $i$-th component of the eigenvector corresponding to the eigenvalue $\lambda_\nu$ as $\psi_i^{(\nu)}$, and we employ the normalisation condition $\sum_i \vert \psi_i^{(\nu)} \vert^2 = 1$. We can then define the global density of states and the eigenvector distribution (at a fixed location $\omega$ in the spectrum)
\begin{align}
\rho(\omega) &= \frac{1}{N} \left\langle \sum_\nu \delta(\omega - \lambda_\nu) \right\rangle, \nonumber \\
P_\psi(y) &= \frac{1}{N^2 \rho(\omega)}\left\langle\sum_{i;\nu}  \delta(\omega - \lambda_\nu)\delta(y - N \vert \psi_i^{(\nu)}\vert^2) \right\rangle ,
\end{align}
where we suppress the dependence of $P_\psi(y)$ on $\omega$ for notational convenience. 

For extended eigenvectors, all of the eigenvector components have roughly the same magnitude of $\vert\psi_i^{(\nu)} \vert^2 \sim 1/N$. For example, the distribution of eigenvector components in the Gaussian orthogonal ensemble (GOE) takes the Porter-Thomas form $P_\psi(y) = e^{-y/2}/\sqrt{2 \pi y}$. In contrast, we demonstrate that the eigenvectors in the extended phase of the weighted ER graphs considered here exhibit many different scales. More precisely, we can define the scaling exponent $\alpha_i^{(\nu)} =- \log(\vert \psi_i^{(\nu)}\vert^2)/\log(N)$, so that $\vert \psi_i^{(\nu)} \vert^2 = N^{-\alpha_i^{(\nu)}}$. One can then characterise the distribution of the different magnitudes of the eigenvector components using
\begin{align}
\Omega(\alpha; \omega) = \frac{1}{N^2 \rho(\omega)} \sum_{i, \nu} \langle \delta(\alpha - \alpha^{(\nu)}_i) \delta(\omega - \lambda_\nu) \rangle .
\end{align}  
\vspace{0.2cm}

One can then define the so-called singularity spectrum $f(\alpha)$ \cite{fyodorov2010multifractality, monthus2017statistical, monthus2011anderson} via the scaling relationship $\Omega(\alpha) = N^{f(\alpha)-1}$ (once again, we suppress the $\omega$ dependence), which we expect to be valid in the limit $N \to \infty$. The singularity spectrum can be related to the distribution of eigenvector components simply via a change of variables
\begin{align}
f(\alpha) = \frac{\log\left[N^2 \vert \psi_i^{(\nu)}\vert^2 P_\psi\left(N \vert \psi_i^{(\nu)}\vert^2\right)\right]}{\log(N)} .\label{ffrompsi} 
\end{align}
We can also define the set of generalised inverse participation ratios \cite{evers2008anderson, fyodorov2010multifractality}
\begin{align}
I_q(\omega) &= \frac{\left\langle \sum_\nu \delta(\omega-\lambda_\nu)\sum_i \vert \psi_i^{(\nu)} \vert^{2q} \right\rangle}{ N \rho(\omega)} \nonumber \\
&= \int_0^N dy \, y^{q} N^{1-q} P_\psi(y) . \label{iprgeneraldef}
\end{align}
Positing the scaling behaviour $I_q \sim N^{-\tau(q)}$ allows the extraction of the set of fractal dimensions $D(q) = \tau(q)/(q-1)$. In the case of the GOE, for which the eigenvectors are fully extended, and we have $D(q) = 1$ identically for $q>1$. For Anderson localised states, in contrast, we have $I_q \sim N^0$. Multifractal states constitute a middle-ground between these two extremes, having $0<D(q)<1$ with $D(q)$ varying as a function of $q$ (i.e. a range of fractal dimensions). The fractal dimensions can be extracted from the singularity spectrum above in the limit $N\to \infty$ via the Legendre transform \cite{da2025spectral, cugliandolo2024multifractal}
\begin{align}
D(q) = -\frac{\max_\alpha \left[f(\alpha) - q \alpha\right]}{q-1}. \label{dfromf}
\end{align}
The quantity at the centre of our analysis, which will permit us to extract the singularity spectrum and the fractal dimensions, is the Local Density of States (LDoS)
\begin{align}
\rho_i(\omega) = \sum_{\nu} \delta(\omega - \lambda_\nu) \vert \psi_i^{(\nu)} \vert^2 . \label{ldosintermsofpsi}
\end{align}
If the distribution $P_\mathrm{LDoS}(x;\omega) = N^{-1} \left\langle\sum_i  \delta(x - \rho_i(\omega)) \right\rangle$ has a power-law tail as a function of $x$, we expect the distribution of eigenvector components $P_\psi(y)$ also to have a power-law tail with the same exponent \cite{da2025spectral, cugliandolo2024multifractal}. Such a power-law tail connotes a non-trivial set of fractal dimensions, and therefore multifractality and non-ergodicity, which we will derive for the ensembles of matrices that satisfy Eq.~(\ref{matstats}).

\section{Local density of states (LDoS)}\label{section:ldos}
A fixture in the analysis of the spectral properties of random matrices is the matrix resolvent (also known as the matrix Green's function)
\begin{align}
G_{ij}(z) = \left(\left[ z \underline{\underline{\id}} - \underline{\underline{J}}\right]^{-1}\right)_{ij} ,
\end{align}
where $z$ is in general complex. The LDoS is obtained from the local resolvent by taking $z$ close to the real line
\begin{align}
\rho_i(\omega) = \frac{1}{\pi} \lim_{\epsilon \to 0^+}\, \mathrm{Im}\,G_{ii}(\omega-i \epsilon), \label{ldosdef}
\end{align}
where $\omega$ and $\epsilon$ here are real. The off-diagonal elements are known to be negligible for large $N$ (see e.g. \cite{biroli2021levy, cizeau1994theory, facoetti2016from, arous2008spectrum}). The diagonal elements can be obtained using the cavity method \cite{rogers2008cavity}. Specifically, we have 
\begin{align}
G_{ii} = \frac{1}{z - \Sigma_i} . \label{cavity}
\end{align}
where we have introduced the self-energy 
\begin{align}
    \Sigma_i=\sum_j J_{ij}^2 G^{(i)}_{jj}\,.
\end{align}
Here, $\underline{\underline{G}}^{(i)}$ is the resolvent matrix of the $N-1 \times N-1$ matrix $\underline{\underline{J}}^{(i)}$, which is formed by removing the row and column $i$ from the original matrix $\underline{\underline{J}}$. These quantities can be determined self-consistently by solving the so-called cavity equations
\begin{align}
    G_{jj}^{(i)} = \frac{1}{z - \Sigma_j^{(i)}} \,,\label{cavity2}
\end{align}
where $\Sigma_j^{(i)}=\Sigma_j-J_{ij}^2G_{ii}^{(j)}$. Eqs.~(\ref{cavity2}) permit a closed solution procedure, and are valid in the large $N$ limit for a locally tree-like network, of which the ER graph is an example. In \cite{rogers2008cavity} the cavity equations were solved numerically for finite single instances of random networks, and analytically in the large-connectivity limit where self-averaging results in considerable simplification. In \cite{bordenave2010resolvent}, Bordenave and Lelarge treat the cavity equations in a distributional sense in the limit of large $N$. It is this latter philosophy that we employ for the remainder of the article: we treat the elements of the resolvent as random variables and seek an approximate solution for their distribution.  Indeed, self-consistent determination of the self-energy distribution in this way is a well-established approach in the theory of localisation \cite{cizeau1994theory, abouchacra1973selfconsistent, biroli2021levy, delapalme2025wishart}.

\section{Perturbative approximation for the self-energy distribution}  \label{section:selfenergydist}   

We now seek to obtain an approximation for the self-energy distribution, which will later help us obtain the LDoS distribution. We do this by performing a perturbative $1/p$ expansion using cavity equations. In the end, we obtain an expression for the self-energy distribution that is accurate up to order $O(p^{-2})$. As such, the approximation that we obtain captures the typical fluctuations of the self-energy. In the subsequent section, we discuss how we may also understand the tails of the LDoS distribution (i.e. the rare fluctuations).

\subsection{Statistics of $\Sigma_i$}
Since the elements $J_{ij}$ and $G_{jj}^{(i)}$ are statistically independent, we may write $\Sigma_i = X_i + i Y_i$ and 
\begin{align}
\langle X_i \rangle &=  \sum_j \left\langle J_{ij}^2 \right\rangle \langle \mathrm{Re}G_{jj}^{(i)}\rangle =\mathrm{Re} \langle G\rangle , \nonumber \\
\langle Y_i \rangle &=  \sum_j \left\langle J_{ij}^2 \right\rangle \langle \mathrm{Im}G_{jj}^{(i)}\rangle =\mathrm{Im} \langle G\rangle, \label{meanself}
\end{align}
where we note that the statistics of the cavity resolvent elements are the same as those of the true resolvent due to the fact that $A_{ij}$ are all independent random variables in the case of the ER graph. We are interested in the fluctuations about this average. Consider the characteristic function of the centred variables $\delta_i^{(X)} = X_i - \mathrm{Re}\langle G \rangle$ and $\delta_i^{(Y)} = Y_i - \mathrm{Im}\langle G \rangle$
\begin{align}
&\mathcal{F}_\Sigma(k_X, k_Y) \equiv \left \langle \exp\left[ i k_X \delta^{(X)}_i + i k_Y \delta_i^{(Y)}\right] \right\rangle \nonumber \\
&= \Bigg \langle \exp\Big[ i  J_{ij}^2( k_X \mathrm{Re}G_{jj} +  k_Y \mathrm{Im}G_{jj} ) \nonumber \\
&\hspace{1cm}- \frac{i}{N}( k_X \mathrm{Re}\langle G\rangle - k_Y \mathrm{Im}\langle G\rangle)\Big] \Bigg\rangle^N . 
\end{align}
In the limit $N\to \infty$, one obtains using the statistics in Eq.~(\ref{matstats})
\begin{align}
\mathcal{F}_\Sigma(k_X, k_Y)\approx&\exp\Bigg[  - \frac{\beta_4}{2!p }\langle\left( k_X \mathrm{Re}\left[G_{jj} \right] +  k_Y \mathrm{Im}\left[G_{jj} \right]\right)^2\rangle  \nonumber \\
&- \frac{i\beta_6}{3! p^2 }\langle\left( k_X \mathrm{Re}\left[G_{jj} \right] +  k_Y \mathrm{Im}\left[G_{jj} \right]\right)^3\rangle\Bigg], \label{genfunctsigma}
\end{align}
where we neglect terms proportional to the higher-order moments of the matrix entries. We comment on the fact that it is necessary to go up to at least $O(p^{-2})$ in our approximation scheme to correctly capture the tail of the LDoS distribution in Section \ref{section:ldosdist}. We also argue that the power-law nature of the tail of the LDoS distribution is not affected by taking into account higher-order terms in $1/p$ in the exponent of Eq.~(\ref{genfunctsigma}) -- only the prefactor would be corrected. 

If we ignore terms $O(p^{-2})$, we see that $\mathcal{F}_\Sigma(k_X, k_Y)$ is just the Fourier transform of a 2-variable Gaussian distribution. A simple way, then, to take into account the higher-order moments systematically is to perform an Edgeworth expansion \cite{wallace1958asymptotic}. That is, we approximate $\exp[- \frac{i\beta_6}{3! p^2 }\langle\left( k_X \mathrm{Re}\left[G_{jj} \right] +  k_Y \mathrm{Im}\left[G_{jj} \right]\right)^3\rangle] \approx 1 - \frac{i\beta_6}{3! p^2 }\langle\left( k_X \mathrm{Re}\left[G_{jj} \right] +  k_Y \mathrm{Im}\left[G_{jj} \right]\right)^3\rangle$, and perform the inverse Fourier transform, obtaining a lengthy but elementary expression for $P_\Sigma(X, Y)$ in terms of the aforementioned Gaussian and its derivatives.

\begin{figure}[t]
	\centering 
	\includegraphics[scale = 0.5]{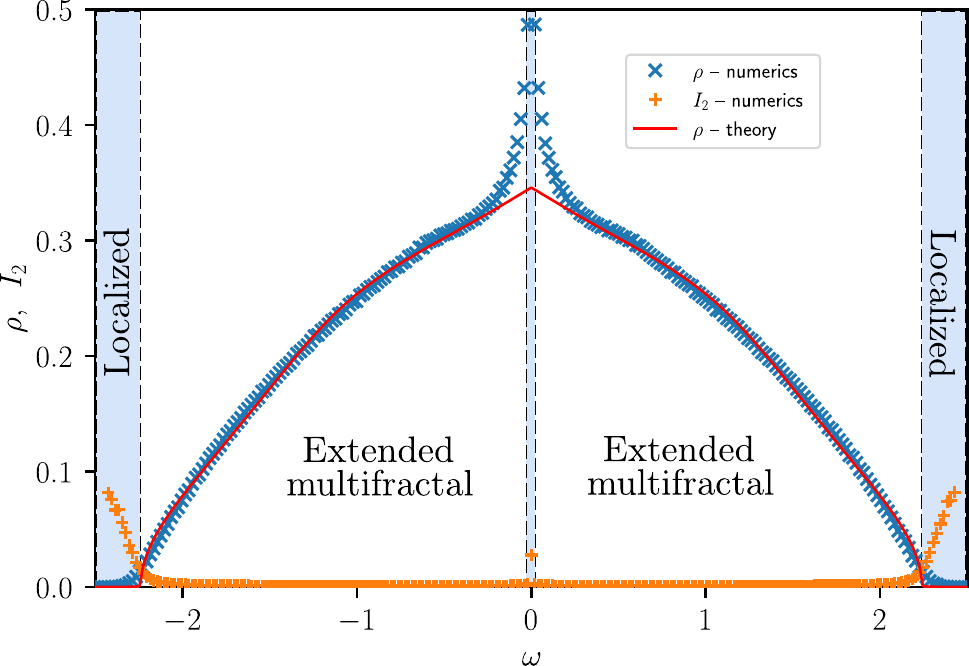}
	\captionsetup{justification=raggedright,singlelinecheck=false, font=small}
	\caption{Global density of states $\rho$ and IPR $I_2$ for a sparse ER graph with uniformly distributed $u_{ij} = W_{ij}^2$. Other weight distributions produce qualitatively similar results. The theory line for $\rho(\omega)$ is found by solving Eq.~(\ref{firstmoment}) numerically with $\beta_4 = 4/3$ and $\beta_6 = 2$ and $p = 5$ and using $\rho = \pi^{-1} \lim_{\epsilon \to 0}\mathrm{Im} \langle G(\omega - i\epsilon) \rangle$. Numerics are the result of diagonalising 500 matrices with size $N = 4000$. }\label{fig:spectrumandipr}
\end{figure}

\subsection{Recursive evaluation of the statistics of $G_{ii}$ using the statistics of $\Sigma_i$ }
Eq.~(\ref{genfunctsigma}) for the characteristic function still contains the statistics of $G_{ii}$ itself, implying the need to solve self-consistently. We approximate the statistics of $G_{ii}$ up to the required order in $p^{-1}$ so that the exponent in Eq.~(\ref{genfunctsigma}) is accurate to $O(p^{-2})$. That is, we must find $\langle G \rangle$ up to order $O(p^{-2})$, find the second moments accurately up to order $p^{-1}$, and the approximation $\langle(\mathrm{Re} G_{ii})^3\rangle\approx [\mathrm{Re}\langle G \rangle ]^3$ (and so on) will suffice for third-order moments.

Let us begin with the mean. This is accomplished by using the definition $G_{ii} = \frac{1}{\omega - X_i - i Y_i}$, leveraging the knowledge that we have already obtained of the cumulants of $X_i$ and $Y_i$. Ultimately, as we show in Appendix~\ref{appendix:avg}, we arrive at the following expression, which is valid up to order $O(p^{-2})$ and can be solved numerically for $\langle G \rangle$
\begin{align}
\langle G \rangle
\approx \frac{1}{\omega - \langle G \rangle} + \frac{\beta_4}{p} \langle G\rangle^5  +  \frac{\beta_6}{p^2}\langle G\rangle^7  + \frac{\beta_4^2}{p^2} \langle G \rangle^9. \label{firstmoment}
\end{align}
This expression was obtained previously using diagrammatic techniques \cite{baron2025path, baron2025classes, baron2022eigenvalue}, and it is verified in Fig. \ref{fig:spectrumandipr}. Using a similar approach, we can also evaluate the higher moments, which enter into the expression for $\mathcal{F}_\Sigma(k_X, K_Y)$ in Eq.~(\ref{genfunctsigma}). We have
\begin{align}
\langle (\mathrm{Re}G_{ii})^2 \rangle &\approx \langle \mathrm{Re}G_{ii} \rangle^2+\frac{\beta_4}{p}\left[\mathrm{Re}\left(\langle G_{ii} \rangle^3 \right) \right]^2, \nonumber \\
\langle (\mathrm{Im}G_{ii})^2 \rangle &\approx \langle \mathrm{Im}G_{ii} \rangle^2+ \frac{\beta_4}{p}\left[\mathrm{Im}\left(\langle G_{ii} \rangle^3 \right) \right]^2  , \nonumber \\
\langle \mathrm{Re}G_{ii} \, \mathrm{Im} G_{ii}\rangle &\approx \langle \mathrm{Re}G_{ii} \rangle\langle \mathrm{Im}G_{ii} \rangle\nonumber \\
&+ \frac{\beta_4}{p}\mathrm{Re}\left(\langle G_{ii} \rangle^3 \right)  \mathrm{Im}\left(\langle G_{ii} \rangle^3 \right) . \label{secondmoments}
\end{align}
Again, these expressions can also be derived using the diagrammatic formalism \cite{baron2025classes}. We note that we need only to evaluate these moments to order $O(p^{-1})$ in order for the neglected contribution in the exponent of Eq.~(\ref{genfunctsigma}) to be of order $O(p^{-3})$. Having already found $\langle G\rangle$, we thus now have the requisite approximation for the second moments. Finally, we may also approximate 
\begin{align}
\langle\left( k_X \mathrm{Re}\left[G_{jj} \right] +  k_Y \mathrm{Im}\left[G_{jj} \right]\right)^3\rangle \approx \left( k_X \mathrm{Re} \langle G \rangle +  k_Y \mathrm{Im}\langle G \rangle \right)^3, \label{thirdmoments}
\end{align}
in Eq.~(\ref{genfunctsigma}). We thus have all the information that we need to evaluate the distribution of the self-energy. Namely, we insert Eqs.~(\ref{firstmoment}), (\ref{secondmoments}) and (\ref{thirdmoments}) into Eq.~(\ref{genfunctsigma}) to obtain the generating function accurate to order $O(p^{-2})$. We then use the Edeworth approximation scheme to invert the Fourier transform [as discussed after Eq.~(\ref{genfunctsigma})] and therefore obtain $P_\Sigma(X, Y)$. 

In summary, by truncating the approximation scheme of the moments at a certain order in $1/p$, we can close the set of equations, and we can express the higher moments of $G_{ii}$ in terms of the mean, for which we obtain a closed-form expression. As a result, we can evaluate the probability distribution of the self-energy to a specified accuracy. 
\begin{figure}[t]
\centering 
\includegraphics[scale = 0.5]{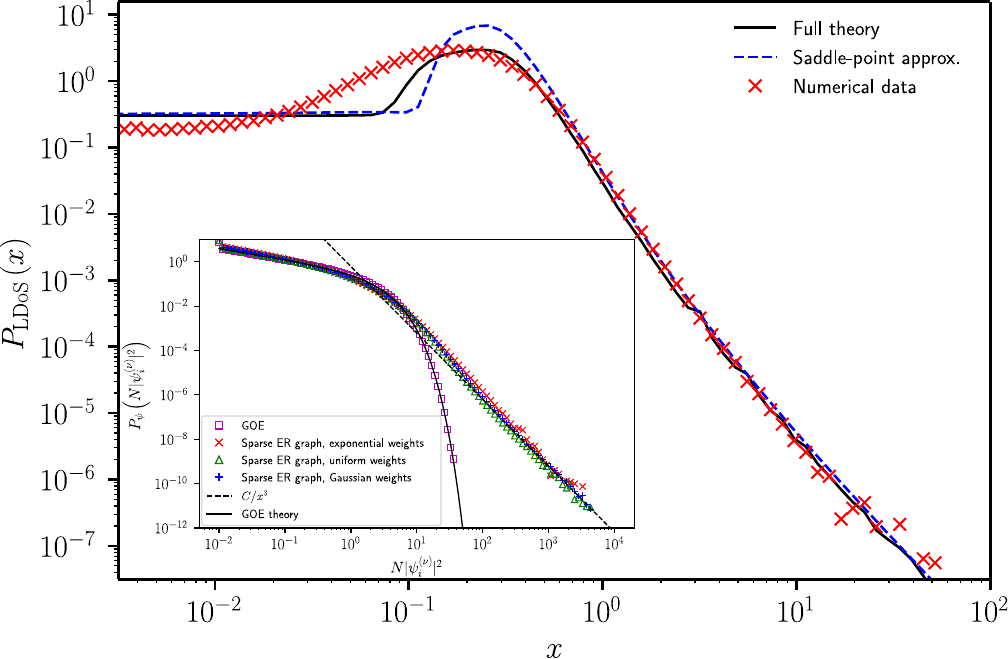}
\captionsetup{justification=raggedright,singlelinecheck=false, font=small}
\caption{Local density of states distribution for the ER graph adjacency matrix with uniformly distributed $u_{ij} = W_{ij}^2$ at $\omega = 0.4$. Comparison is made with Eq.~(\ref{fullsol}) (solid black line) and the sum of Eqs.~(\ref{asymp}) and (\ref{smallxapprox}) (dashed blue line). (Inset) The distribution $P_\psi(y)$ for the three weight distributions described in Section \ref{section:model}, The theory line $P_\psi(y) = B/y^3$ is shown as a dashed line (with the constant $B$ numerically fitted). Parameters: $p = 3.4$, $\omega = 0.4$, $N = 10^4$, and results are averaged over 100 trials. }\label{fig:ldosdist}
\end{figure}
\section{LDoS distribution and power-law tail}\label{section:ldosdist}

\subsection{General LDoS distribution expression}
Now that we have approximated the distribution of the self-energy, we may evaluate the object that is most central to our considerations, the distribution of the LDoS. Since the LDoS can be written in terms of the self-energy using Eqs.~(\ref{ldosdef}) and (\ref{cavity}), 
\begin{align}
\rho_i = \frac{1}{\pi}\frac{Y_i}{(X_i - \omega)^2 + Y_i^2},\label{imgir}
\end{align}
we can write the LDoS distribution $P_\mathrm{LDoS}(\rho_i =x)$ in terms of the distribution of the self-energy as follows. Using a simple change of variables, we have
\begin{widetext}

\begin{align}
P_\mathrm{LDoS}(x) &= \int_{\omega-\frac{1}{2\pi x}}^{\omega+\frac{1}{2\pi x}} dX  \, \left[\frac{\pi(Y_+^2 + (X-\omega)^2)^2}{Y_+^2 - (X-\omega)^2}P_\Sigma\left(X, Y_+(x,X)\right) - \frac{\pi(Y_-^2 + (X-\omega)^2)^2}{Y_-^2 - (X-\omega)^2}P_\Sigma\left(X, Y_-(x,X)\right)\right] ,  \label{fullsol}
\end{align}
\end{widetext}
where $Y_\pm(x, X) = \frac{1 \pm \sqrt{1 - 4\pi^2 x^2(\omega-X)^2}}{2 x \pi}$ and the integration range ensures that $Y_\pm(x,X)$ is real. The expression in Eq.~(\ref{fullsol}) is the most general form for the LDoS distribution, and it can be evaluated numerically by first computing an analytical expression for the distribution $P_\Sigma\left(X, Y_-(x,X)\right)$ using the Edgeworth approach, as discussed in the previous section. One thus obtains the solid black line in Fig. \ref{fig:ldosdist}, which demonstrates good agreement with numerics, especially in the tail region where $x$ is large. 

\subsection{Asymptotic behaviour}
We see from Fig. \ref{fig:ldosdist} that there appears to be a power-law behaviour in the tail region of the curve $P_\mathrm{LDoS}(x)$. We can confirm this and extract the power-law exponent by making a large-$x$ approximation of the integral in Eq.~(\ref{fullsol}) using Laplace's method. For large $x$, one finds that the value of $X$ at which the integrand in Eq.~(\ref{fullsol}) is maximum is $X^\star = \omega + O(x^{-2})$. This is to be expected since, in order for the right-hand side of Eq.~(\ref{imgir}) to be large, we must have $X \approx \omega$ and small $Y$. One thus arrives at [see Appendix \ref{appendix:laplace} for details of the calculation]
\begin{align}
P_\mathrm{LDoS}(x) &\approx \frac{1}{(\sqrt{\pi}x)^3}P_\Sigma\left(X = \omega, Y = \frac{1}{\pi x}\right). \label{asymp}
\end{align}
We therefore see that the distribution of the LDoS has a power-law tail with exponent $3$ as long as $P_\Sigma\left(X = \omega, Y = \frac{1}{\pi x}\right)$ tends to a constant for large $x$. We expect this to be the case as long as the distribution of $W_{ij}^2$ (and consequently $Y_i$) has a non-zero and finite probability mass at small values. We emphasise that the value of $3$ here is not a result of our approximation scheme for $P_\Sigma(X,Y)$ and is instead a generic property of the weighted ER graph ensembles. This is evident from the generality of the reasoning in Appendix \ref{appendix:laplace}, where we do not appeal to a specific form for $P_\Sigma(X,Y)$. 

A symmetry that applies to the LDoS distribution in many cases is \cite{fyodorov2004statistics}
\begin{align}
P_\mathrm{LDoS}(x) =  (x/\rho)^{-3}P_{\mathrm{LDoS}}(\rho^2 x^{-1}). \label{symmetry}
\end{align}
This symmetry arises from an analysis of the non-linear sigma model \cite{fyodorov2013universal, mirlin2006exact}, and has been verified to apply in a broad range of circumstances \cite{de2014anderson, cugliandolo2024multifractal, evers2008anderson}. In the present case, the symmetry would imply that $P_{\mathrm{LDoS}}(x) \sim x^0$ for small $x$. Indeed, we can show that this symmetry is satisfied asymptotically for small $x$ by once again performing a saddle-point approximation. This procedure is also outlined in Appendix \ref{appendix:laplace}. We discuss how the symmetry in Eq.~(\ref{symmetry}) manifests in the eigenvector statistics in Section \ref{section:singularityspectrum}.

The sum of the two asymptotic predictions (small and large $x$) is compared against both the numerical integration of Eq.~(\ref{fullsol}) and numerically generated random matrices in Fig. \ref{fig:ldosdist}. We indeed find the same power-law exponent in the tail for a range of random matrix ensembles, as should be the case, given the generality of our reasoning. We note that the exponent of 3 is close to the values found numerically (ranging between $2.2$ and $3.2$) in Ref. \cite{cugliandolo2024multifractal} for weighted sparse ER graphs. However, we note that the weight distribution in this work did not satisfy $0<\lim_{u \to 0}P(u)<\infty$ (where, again, $u_{ij} = W_{ij}^2$), which could account for the small discrepancy.

\subsection{Onset of power-law behaviour and cross-over to ergodicity}
We can also obtain a rule-of-thumb estimate for the value of $x$ at which the onset of the power-law behaviour occurs. Expanding Eq.~(\ref{asymp}) in $x^{-1}$, we obtain
\begin{align}
P_\mathrm{LDoS}(x) &\approx \frac{P_\Sigma(\omega,0)}{(\sqrt{\pi}x)^3 } \left[ 1+ \frac{1}{\pi^2 x} \frac{\omega\, p^2}{4 \beta_4^2  \rho \mathrm{Re}\langle G \rangle \vert \langle G \rangle \vert^4 }  + \cdots \right].
\end{align}
We enter the power law regime when the second term in the square brackets becomes negligible in comparison to the first. This crossover occurs when $x \approx p^2/[2 \pi^2\beta_4^2 \rho(\omega)]$, where we use that $\mathrm{Re}\langle G \rangle \approx \omega/2$ and $\vert \langle G\rangle\vert \approx 1$ to leading order in $1/p$. That is, the power-law tail becomes suppressed as the average degree of the ER graph is increased, also in agreement with the numerical findings of Ref. \cite{cugliandolo2024multifractal} that found a similar gradual crossover from non-ergodic to ergodic behaviour.

\subsection{Breakdown of the $x^{-3}$ behaviour}\label{section:breakdown}
While the $x^{-3}$ behaviour in the LDoS distribution is common across the weighted ER graph ensembles that we study in this work, it is not a completely general phenomenon. In particular, as can be seen from Eq.~(\ref{asymp}), the asymptotic behaviour of $P_\mathrm{LDoS}(x)$ for large $x$ is related to the asymptotic behaviour of $P_\Sigma(X,Y)$ for small $Y$. That is, we only obtain a $1/x^3$ tail of the LDoS distribution if the self-energy distribution tends to a constant at small values its argument. It is for this reason that we chose weights whose squares $u_{ij} = W_{ij}^2$ had distributions $P(u)$ with non-zero probability mass at zero. 

Related to this, we noted earlier that it was necessary to go to second order in $1/p$ in our approximation scheme to understand the tail of the distribution $P_\mathrm{LDoS}(x)$. The reason for this is that if we ignore the $O(1/p)$ terms in Eq.~(\ref{secondmoments}) and the other contribution to the $O(1/p^2)$ in Eq.~(\ref{genfunctsigma}), we find that $\mathcal{F}_\Sigma$ is simply the generating function of two linearly dependent Gaussian variables. That is, we could effectively write $\delta^{(X)} = z\mathrm{Re}\langle G \rangle $ and $\delta^{(Y)} = z\mathrm{Im}\langle G \rangle $, where $z$ is a centred Gaussian random variable with variance $\beta_4/p$. The linear dependence of $X$ and $Y$ would mean that, according to Eq.~(\ref{imgir}), $\rho_i$ would have a cut-off at $(\mathrm{Re}\langle G \rangle  + \vert \langle G \rangle \vert)/(2 \omega\mathrm{Im}\langle G \rangle )$, which we do not observe. Including the $1/p$ correction in Eq.~(\ref{secondmoments}), in particular, breaks the linear dependence of $X$ and $Y$, and ultimately gives rise to the tail of $P_\mathrm{LDoS}(x)$. 

Another way in which the $1/x^3$ behaviour can be suppressed is by moving $\omega$ to the edge of the continuous part of the eigenvalue spectrum, whereupon we observe a different behaviour. As has been found previously in the context of Anderson localisation \cite{mirlin1994distribution, kravtsov2018nonergodic, tikhonov2019statistics, mirlin1994statistical, biroli2022critical}, we find a power-law behaviour of $P_\mathrm{LDoS}(x)\sim x^{-3/2}$ close to the mobility edge. This is discussed further in Section \ref{section:edge}. This same power-law exponent is carried over to the localised phase, where the spectrum is no longer continuous but is instead point-like, and a non-trivial dependence of the LDoS distribution on the regulariser $\epsilon$ emerges \cite{biroli2018delocalization, biroli2021levy}. 

Finally, one also obtains different asymptotic behaviour in the special case where $\omega = 0$ precisely. This is because the local resolvent in this case is purely imaginary. As such, we may write simply $\rho_i = (\pi Y_i)^{-1}$. If we define $P_Y(Y)$ to be the distribution of $Y$ in this case, using that $X = 0$, one thus obtains 
\begin{align}
P_\mathrm{LDoS}(x;\omega = 0) = \frac{1}{\pi x^2} P_Y\left(\frac{1}{\pi x}\right) .
\end{align}
We discuss why this means that states with $\omega = 0$ are localised in Section \ref{section:edge}. Such localised states have been identified in complex network adjacency matrices previously in Ref. \cite{tapias2023multifractality}.

\section{Singularity spectrum}\label{section:singularityspectrum}
\begin{figure}[t]
\centering 
\includegraphics[scale = 0.5]{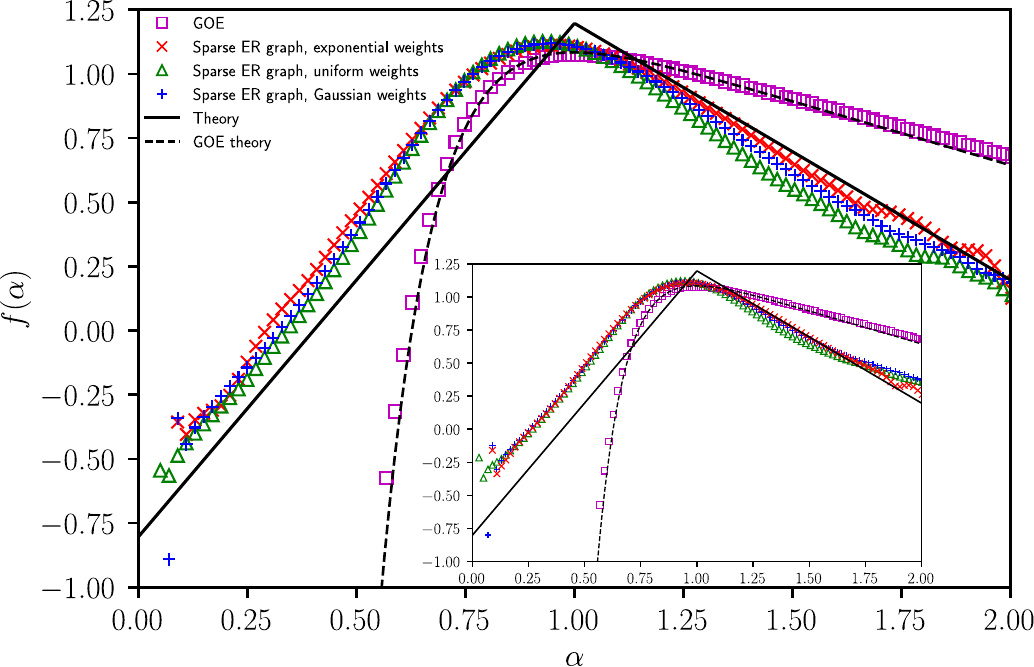}
\captionsetup{justification=raggedright,singlelinecheck=false, font=small}
\caption{Singularity spectrum for the three weight distributions mentioned in Section \ref{section:model} at $\omega = 0.4$, with $p = 3.4$ and $N = 10^4$ averaged over $100$ trials. GOE is also given for contrast. The solid black line is the result in Eq.~(\ref{singularityspectrum}). (Inset) The case of non-zero diagonal disorder with $W = 0.3$. The singularity spectrum remains unchanged when $W$ (the amount of diagonal disorder) is sufficiently small.  }\label{fig:singularity}
\end{figure}

Given the relationship between the LDoS and the eigenvectors shown in Eq.~(\ref{ldosintermsofpsi}), we surmise that the eigenvector distribution itself must also have a power-law tail with the same exponent \cite{da2025spectral, cugliandolo2024multifractal}. Indeed, this is verified in the inset of Fig. \ref{fig:ldosdist}. Our characterisation of the $1/x^3$ tail therefore allows us to make some deductions about the singularity spectrum $f(\alpha)$. 

The power law tail, which occurs for large values of $y = N\vert \psi_i^{(\nu)}\vert^2 = N^{1-\alpha^{(\nu)}_i}$, corresponds to small values of $\alpha$. Substituting $P_\psi(y) =B  y^{-3}$ into Eq.~(\ref{ffrompsi}), one obtains
\begin{align}
f(\alpha) = 2\alpha -1,
\end{align}
noting that the contribution from the constant $B$ vanishes in the limit $N\to\infty$. To understand the behaviour of the eigenvectors on smaller scales, previous studies have exploited the symmetry $f(1+\alpha) = f(1-\alpha) + \alpha$ \cite{de2014anderson, cugliandolo2024multifractal}. This symmetry derives from the previously mentioned symmetry $P_\mathrm{LDoS}(x) = (x/\rho)^{-3} P_\mathrm{LDoS}(\rho^2 x^{-1})$ [see the discussion around Eq.~(\ref{asymp})]. If the symmetry were to hold for the eigenvectors, we would expect $f(\alpha) = 2-\alpha$ for $\alpha\gtrsim 1$ and thus
\begin{align}
f(\alpha) \approx \begin{cases}
2\alpha -1 \hspace{2cm}\mathrm{for}\hspace{2cm} \alpha \ll 1, \\
2-\alpha \hspace{1.45cm}\mathrm{for}\hspace{2cm} \alpha \gg 1 .
\end{cases}\label{singularityspectrum}
\end{align}
With that being said, what we notice in numerics is behaviour more akin to GOE statistics for small $y = N\vert \psi_i^{(\nu)}\vert^2$ (see the inset of Fig. \ref{fig:ldosdist}). However, as was pointed out in Ref. \cite{de2014anderson}, this is because rapid GOE/Porter-Thomas fluctuations dominate when $y = N\vert \psi_i^{(\nu)}\vert^2$ is small. To obtain the slowly varying envelope that characterises multifractal eigenstates, one must separate the fast fluctuation via a deconvolution procedure. That is, one assumes $\log y = \log y_\mathrm{GOE} + \log y_\mathrm{env}$, where $y_\mathrm{GOE}$ obeys Porter-Thomas statistics. Following the procedure detailed in Ref. \cite{de2014anderson}, we may therefore extract the distribution $P_\mathrm{env}(y_\mathrm{env})$ numerically, which yields the singularity spectrum via Eq.~(\ref{ffrompsi}). Indeed, after the deconvolution procedure, the singularity spectrumseems to obey the symmetry $f(1+\alpha) = f(1-\alpha) + \alpha$ and therefore Eq.~(\ref{singularityspectrum}), as is shown in Fig.~\ref{fig:singularity}.

\begin{figure}[t]
\centering 
\includegraphics[scale = 0.5]{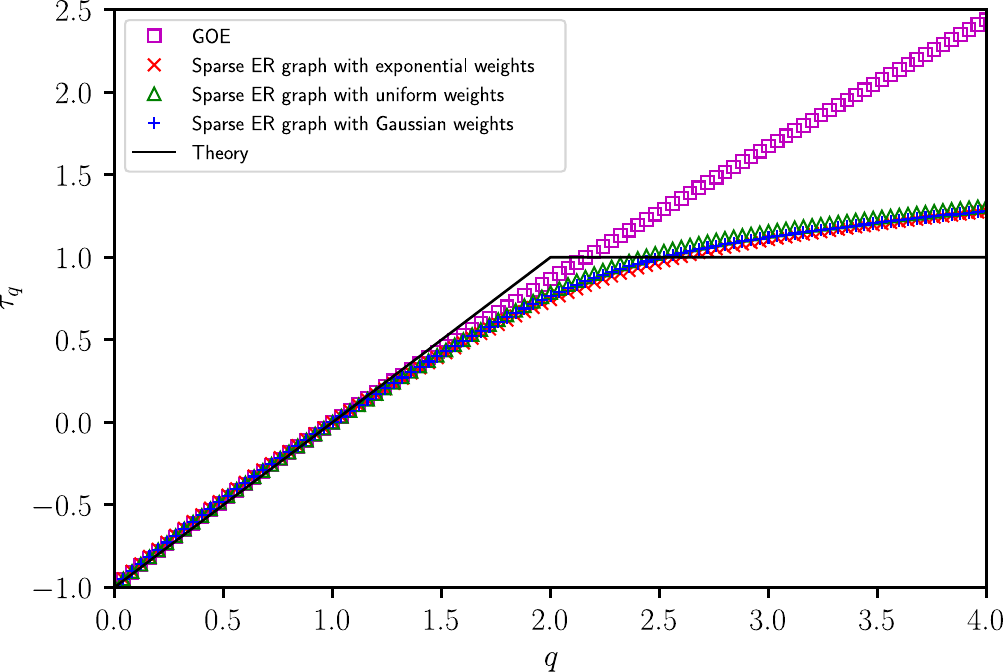}
\captionsetup{justification=raggedright,singlelinecheck=false, font=small}
\caption{The scaling exponent $\tau_q = D(q) (q-1)$ for the same ensembles and parameters as in Fig. \ref{fig:singularity}. The solid line is given by the result in Eq.~(\ref{fractaldimensions}). }\label{fig:multifractality}
\end{figure}

\section{Weak multifractality and non-ergodicity}\label{section:nonergodicity}
Given the behaviour of the singularity spectrum, we may now extract the fractal dimensions using Eq.~(\ref{dfromf}), and therefore gain insight into the ergodicity (or lack thereof) of the eigenvectors in the extended phase. 

The precise behaviour of $f(\alpha)$ for $\alpha>1$ turns out to be unimportant for constructing $\tau(q)$ using the Legendre transform. All that matters is that the behaviour $f(\alpha) = 2 \alpha - 1$ stops at $\alpha \approx 1$, where the function $f(\alpha)$ is maximum, and beyond which $f(\alpha)$ decreases. We thus have
\begin{align}\label{fractaldimensions}
D(q) = \begin{cases}
1 \hspace{2cm}\mathrm{for}\hspace{2cm} 0\leq q\leq2, \\
1/(q-1) \hspace{0.8cm}\mathrm{for}\hspace{2cm} 2\leq q .
\end{cases}
\end{align}
The degenerate nature of the fractal dimensions here is the signature of multifractality. We see however that the multifractality here manifests only in the higher moments of the eigenvectors. This is then `weak' multifractality, in the sense used by Evers and Mirlin \cite{evers2008anderson}, in that the behaviour does not deviate much from the typical `metallic' kind for a certain range of $q$. This is in contrast with stronger forms of multifractality, as seen in the modified Rosenzweig-Porter models \cite{biroli2021levy, khaymovich2021dynamical, khaymovich2020fragile}, the Anderson model on the Bethe lattice \cite{biroli2018delocalization, monthus2011anderson}, and at the Anderson critical point in finite dimensions \cite{evers2008anderson, mirlin2000multifractality}. As was also noted in Ref. \cite{cugliandolo2024multifractal}, this kind of behaviour is similar to that of generalised $\beta$ ensembles \cite{das2024robust, das2023absense}, for which the fractal dimension differs from $D(q) = 1$ only for $q>1$.

\begin{figure}[t]
\centering 
\includegraphics[scale = 0.5]{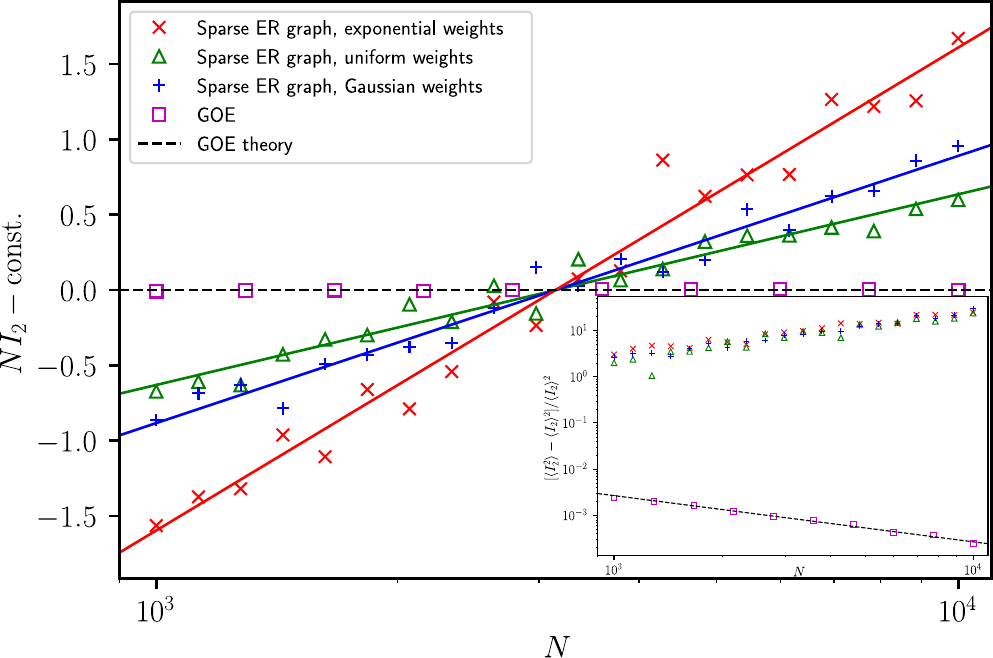}
\captionsetup{justification=raggedright,singlelinecheck=false, font=small}
\caption{Logarithmic scaling of the IPR with $N$ for system parameters and ensembles given in Fig. \ref{fig:singularity}. We have shifted the data so that each set of points passes through the origin. We show lines of best fit as guide to the eye. Since the constant $B$ varies between ensembles (where $P_\psi(y) = B/y^3$ asymptotically), the gradients of the best fit lines in the above plot also vary. (Inset) The relative fluctuations of the IPR $[\langle I_2^2\rangle - \langle I_2\rangle^2]/\langle I_2\rangle^2$ grow with $N$ while they vanish for the GOE. }\label{fig:logn}
\end{figure}

While sometimes non-ergodicity and multifractality are considered as almost synonymous, non-ergodicity can be considered somewhat distinct from the (multi)fractality of the eigenvectors/states. Non-ergoditicty has been defined in several ways in different contexts, leading to various (not always commensurate) criteria for its presence \cite{luitz2017ergodic}. Criteria for non-ergodicity include, for example, a deviation from Wigner-Dyson eigenvalue statistics \cite{luitz2017ergodic}, violation of the eigenstate thermalisation hypothesis \cite{pal2010many, luitz2016long}, a lack of convergence of $\sum_i \vert \psi_i^{(\nu)}\vert^{2q}$ to the ensemble average value \cite{de2014anderson}, or indeed a deviation of the fractal dimensions from the values expected of fully extended states. 

We have already demonstrated that the fractal dimensions deviate from $D_q = 1$ when $q>2$. However, two of the scaling dimensions, $D_1$ and $D_2$ have particular physical relevance for quantifying non-ergodicity \cite{luitz2017ergodic, torres2017extended, biroli2018delocalization, altshuler2016nonergodic}, as we discuss below. We here use $D_1$ and $D_2$ to understand in what sense the eigenvectors of the random matrix ensembles under present consideration are non-ergodic. 

A widely used measure of ergodicity/localisation is the inverse participation ratio (IPR), $I_2$. The reason for this is that the IPR is related to the infinite time average of the survival probability of a quantum state \cite{luitz2017ergodic, torres2017extended}, and so $1/I_2 = N^{D_2}$ is an obvious quantifier of the number of sites occupied by that state. One also has that $I_2 \to 0$ for extended states, and $I_2 \sim N^0$ for localised states, and hence a non-ergodic extended state would be one for which $I_2 \to 0$, but $1/(NI_2) \to 0$ as $N \to \infty$, i.e. one in which only a subextensive number of sites are accessed. Here, we find that (as is verified in Fig. \ref{fig:logn})
\begin{align}
I_2 = \int_0^N dy \, y^2 N^{-1} P_\psi(y) \sim \frac{\ln N}{N}, \label{iprfromppsi}
\end{align}
meaning that states indeed occupy a vanishing fraction of the sites [with $1/(NI_2) \to 0$]. However, we also have $D_2 \approx 1- \ln \ln N/\ln N \to 1$ in the thermodynamic limit. So, we have a somewhat special situation. One can say that, by the above criterion, we have non-ergodicity, but only a weak kind in which the relevant scaling dimension tends to unity for $N \to \infty$ (albeit very slowly). We further verify in the inset of Fig. \ref{fig:logn} that the relative fluctuations (another indicator of non-ergodicity \cite{de2014anderson}) of the IPR are non-vanishing, and in fact grow with $N$, while in contrast the IPR fluctuations decrease with $N$ for the fully extended states of the GOE.

Let us now turn to another common quantifier of non-ergodicity, $D_1$, which is related to the Shannon entropy via $S = D_1 \ln N$. Following Kravtsov et al \cite{kravtsov2018nonergodic, de2013support}, the `support set', of which $D_1$ is the scaling dimension, can be defined as the least number of components required to satisfy the normalisation constraint up to some prescribed accuracy $\epsilon$ (see Appendix \ref{appendix:support_set} for further details). One can show that the support set of the eigenvectors is $M_\epsilon = A_\epsilon N^{D_1}$. The criterion in Ref. \cite{kravtsov2018nonergodic} regards eigenvectors with $M_\epsilon \sim N^0$ as localised and those with $M_\epsilon \to \infty$ as extended (as $N \to \infty$), but if $M_\epsilon/N \to 0$ then the state is non-ergodic. One immediately sees that if $0<D_1<1$, then the state is non-ergodic and extended by this definition. 

Since $D_1= 1$ in the present case [see Eq.~(\ref{fractaldimensions})], the support set is extensive (this was also pointed out in the context of sparse random graphs in Ref. \cite{cugliandolo2024multifractal}). We can confirm this by computing the support set explicitly according to the prescription of Ref. \cite{kravtsov2018nonergodic}, which we do in Appendix \ref{appendix:support_set}. We find that the eigenvector support set is indeed extensive $M_\epsilon \approx A_\epsilon N$, where $A_\epsilon \approx  (1-\epsilon)^2/(2 B)$, where $B$ is same constant that appears in the power law $P_\psi(y) = B/y^3$. This is in contrast to the non-ergodic extended states in models such as the L\'evy-Rosenzweig-Porter ensemble \cite{biroli2021levy} or the Anderson model on the Bethe lattice \cite{kravtsov2018nonergodic, biroli2018delocalization}, where $D_1<1$. We also note that the dimension $D_1$ is related to the peak of the LDoS distribution via $e^{\langle \ln \rho_i \rangle} \sim N^{D_1-1}$ \cite{biroli2018delocalization}. The peak of the LDoS distribution in our case is smaller than the mean, but not parametrically so, and this is expressed accordingly by the fractal dimension $D_1 = 1$.

In summary, although according to the criteria involving $I_2$ and its fluctuations the eigenvectors here are in some sense non-ergodic, and although the scaling dimensions for $q>2$ deviate from $D_q = 1$, this is a weak effect in comparison to some other random matrix ensembles. 

\section{Spectral edge, localised phase and varying diagonal elements}\label{section:edge}
Having identified and understood the nature of the non-ergodic extended phase in the weighted ER graph ensembles, we examine what happens at the edge of the continuous part of the eigenvalue spectrum, where this behaviour terminates, and we see the onset of $I_2 \sim N^0$ (see Fig. \ref{fig:spectrumandipr}). As has been shown in the context of sparse graphs \cite{biroli1999single, semerjian2002sparse, valigi2025eigenvalue}, it is local network `defects' (i.e. well-connected hubs or heavily weighted edges) that give rise to these eigenvalues outside the continuous part of the eigenvalue spectrum. One can indeed show that the corresponding eigenvectors are exponentially localised about such defects.  

As discussed in Appendix \ref{appendix:edge}, we demonstrate that at the spectral edge the LDoS distribution no longer obeys the power law in Eq.~(\ref{asymp}). Instead, a different power-law behaviour emerges, which is characteristic of the transition to a localised phase, and has a power-law exponent of $3/2$, as has also been found previously in the context of Anderson localisation \cite{mirlin1994distribution, kravtsov2018nonergodic, tikhonov2019statistics, mirlin1994statistical, biroli2022critical}. This same power law behaviour is carried over into the localised phase, but a dependence on the regulariser emerges \cite{biroli2018delocalization, biroli2021levy}. 

On the subject of localisation, we may also ask what happens when we include non-zero diagonal elements $J_{ii}$, for instance drawn from a uniform distribution on the interval $[-W, W]$, as is common in the classic Anderson tight-binding model \cite{monthus2011anderson, evers2008anderson, biroli2010anderson}. For values of $\omega$ and $W$ such that $\vert\omega \pm W\vert<\omega_c$, where $\omega_c = 2 +O(p^{-1})$ is the spectral edge \cite{baron2025classes, baron2025path, rodgers1988density}, we expect that the power-law tail of the LDoS should be preserved. 

The reason for this can be seen by the following simple argument. Suppose we consider the subset of sites with $J_{ii} = E$. The LDoS statistics of these sites would simply be given by the LDoS statistics of the model without disorder at $\omega-E$. In the tail region, we would thus have
\begin{align}
P_\mathrm{LDoS}(x;\omega, E) = P_\mathrm{LDoS}(x;\omega-E, 0) \approx \frac{A(\omega-E)}{x^3} ,
\end{align}
with $A(\cdot)$ being a function that is independent of $x$. Therefore, integrating over the possible values of $E$, we obtain the same power-law behaviour in the tail. If however $W$ is large enough so that $\omega-E$ can be outside the continuous spectrum, this reasoning no longer applies, since the LDoS distribution no longer maintains the same tail behaviour. We instead expect that the LDoS distribution will pick up a dependence on the regulariser, the $3/2$ power-law behaviour will dominate over the quicker power-law in the bulk, and states will become localised in the usual way. That the eigenvector components maintain the same multifractal statistics in the extended phase is verified in the inset of Fig. \ref{fig:singularity}.

Finally, as we remarked earlier in Section \ref{section:breakdown}, the states with precisely $\omega = 0$ are also localised. One can see this from the asymptotic behaviour $P_\mathrm{LDoS}(x) \sim x^{-2}$ for large $x$, which connotes $P_\psi(y) \sim y^{-2}$ for large $y$. We can therefore calculate the IPR using Eq.~(\ref{iprfromppsi}), and we obtain $I_2 \sim O(N^0)$.  

\section{Summary and discussion}\label{section:summary}
In this work, we have studied the multifractal properties of the eigenvectors of certain weighted Erd\H{o}s-R\'enyi graphs. By using a simple cavity-like approach \cite{rogers2008cavity, susca2021cavity}, we were able to use a systematic approximation scheme to quantify the distribution of the local density of states. Upon finding that the tail of this distribution exhibited a power-law behaviour with exponent $3$, we were able also to understand the statistics of the eigenvectors.

We found that, as a consequence, the eigenvectors were weakly multifractal. We also showed that the inverse participation ratio of the eigenvectors scaled as $\ln N/N$, meaning that the eigenvectors were extended and weakly non-ergodic. In other words, the number of sites occupied by the eigenvectors (as judged by the IPR) diverged in the $N\to \infty$ limit, but the occupied fraction of the total number of sites was vanishing, and scaled as $1/\ln N$. This is in contrast to stronger kinds of non-ergodicity, where the occupied number of sites goes as $\sim N^{D_2}$ with $0<D_2<1$, connoting a non-trivial fractional scaling dimension. Only in the higher moments of the eigenvector components do we see a deviation from fully ergodic behaviour that affects the fractal dimensions.

We note that the peculiar weak non-ergodicity and multifractality displayed here has a fingerprint in the eigenvalue correlations also. Random matrix models exhibiting strong multifractality are associated with `minibands' of $N^{D_1}$ hybridised eigenvalues in the non-ergodic extended phase \cite{biroli2021levy, khaymovich2020fragile}, and a deviation from Wigner-Dyson statistics of the microscopic eigenvalue correlations. In contrast, we have $D_1 = 1$ in our case, and as was shown in Ref. \cite{baron2025classes}, the microscopic eigenvalue statistics remain of the Wigner-Dyson type. However, there are instead long-range eigenvalue correlations, which are not present for ensembles like the GOE. Such long-range correlations still give rise to a non-zero level compressibility $0<\chi<1$, marking a half-way point between the usual Wigner-Dyson result $\chi \to 0$ and the value for Poisson statistics $\chi \approx 1$. This somewhat mirrors the aforementioned unusual `in-between' nature of the IPR found here, in that there is a marked deviation from the usual fully-extended eigenvector behaviour, but not one that is detected in the fractal dimension $D_1$.

Physically, the origin of the multifractal extended states is a consequence of the combination of two model aspects: the sparsity of the ER graph and the weights being allowed to come arbitrarily close to zero. As was also observed in Ref. \cite{cugliandolo2024multifractal}, this makes it possible for `hidden' isolated clusters to form (i.e. distinct well-connected clusters with only weak connections between them), even when we consider only the giant component of the ER graph. What is only elucidated through analysis however is the power-law exponent of $3$ that we have found here in the case where $0<\lim_{u \to 0}P(u)<\infty$. We anticipate that a range of exponents (different from $3$) could be found for other weight distributions, and this remains a subject of our investigation. 

A natural extension to the analysis performed here would be to examine the LDoS distribution of other random matrix ensembles using the cavity approach, particularly those exhibiting strong multifractality. The use of a scheme whereby one determines the self-energy distribution self-consistently, in the context of localisation phenomena, is not unprecedented \cite{cizeau1994theory, abouchacra1973selfconsistent, biroli2021levy}, but certainly there is opportunity for a more thorough evaluation of the LDoS distribution in many important random matrix models. However, we note that one advantage that was available to us here was the fact that the generating function of the self-energy was independent of $N$, and was amenable to a perturbative treatment, which allowed closure of the self-consistent equations. In models such as the L\'evy-Rosenzweig-Porter ensemble \cite{biroli2021levy, safonova2025spectral}, the $N$ dependence and statistics of the self-energy are somewhat more complicated, and may require non-trivial amendments to the analytical procedure, but this remains an intriguing avenue for future inquiry.

\acknowledgements
The authors would like to thank Boris Altshuler, Jean-Philippe Bouchaud, Yan Fyodorov, Reimer Kuehn, Dmitry Savin, Marco Tarzia and Davide Venturelli for enlightening discussions. JWB thanks the Leverhulme Trust for support through the Leverhulme Early Career Fellowship scheme. 

\appendix

\section{Evaluation of $\langle G \rangle$ }\label{appendix:avg}
Let us begin with the expression in Eq.~(\ref{cavity}), recalling the definitions $\delta_i^{(X)} = X_i - \mathrm{Re}\langle G \rangle$, $\delta_i^{(Y)} = Y_i - \mathrm{Im}\langle G \rangle$, and $\delta \Sigma_i = \delta_i^{(X)}+ i \delta_i^{(Y)}$. Given that from Eq.~(\ref{genfunctsigma}) we expect higher cumulants to give rise to higher-order terms in $1/p$, we may perform a systematic approximation of $\langle G \rangle$ to the desired order in $1/p$. Proceeding as such, we have
\begin{align}
&\langle G \rangle = \left\langle \frac{1}{\omega - \langle G \rangle - \delta_i^{(X)} - i \delta_i^{(Y)}} \right\rangle  \nonumber \\
&\approx \frac{1}{\omega - \langle G \rangle} + \frac{\langle(\delta \Sigma)^2\rangle}{(\omega - \langle G \rangle)^3} +  \frac{\langle(\delta \Sigma)^3\rangle}{(\omega - \langle G \rangle)^4}  + \frac{\langle(\delta \Sigma)^4\rangle}{(\omega - \langle G \rangle)^5} \nonumber \\
&\approx \frac{1}{\omega - \langle G \rangle} + \frac{\beta_4 \langle G^2\rangle}{p(\omega - \langle G \rangle)^3} +  \frac{\beta_6\langle G^3\rangle}{p^2(\omega - \langle G \rangle)^4}  \nonumber \\
&\,\,\,\,\,\,\,+ \frac{3 \beta_4^2 \langle G^2\rangle^2}{p^2(\omega - \langle G \rangle)^5}, \label{avresintermsofbeta}
\end{align}
where we have kept only the terms that contribute to order $O(p^{-2})$. Now, we wish to write the expression for $\langle G \rangle$ in terms of only itself. To do this, we exploit the fact that our approximation scheme only requires us to approximate $\langle G^2\rangle$ to order $O(p^{-1})$ and $\langle G^3\rangle$ to order $O(p^0)$ in order for the expression in Eq.~(\ref{avresintermsofbeta}) to be valid to order $O(p^{-2})$. This permits us to close the set of equations.

Therefore, for the third moment, we may simply use $\langle G^3 \rangle \approx \langle G \rangle^3$. For the second moment, we have instead
\begin{align}
&\langle G^2 \rangle = \left\langle \left[\frac{1}{\omega - \langle G \rangle - \delta_i^{(X)} - i \delta_i^{(Y)}} \right]^2\right\rangle \nonumber \\
&\approx \left[\frac{1}{\omega - \langle G \rangle} \right]^2 +  \frac{2\beta_4\langle G^2\rangle}{p(\omega - \langle G \rangle)^3} + \frac{\beta_4\langle G^2\rangle}{p(\omega - \langle G \rangle)^4} \nonumber \\
&\approx \langle G \rangle^2 + \frac{\beta_4\langle G^2\rangle}{p(\omega - \langle G \rangle)^4}\approx \langle G \rangle^2 + \frac{\beta_4\langle G\rangle^2}{p(\omega - \langle G \rangle)^4},
\end{align}
where we have used the expression in Eq.~(\ref{avresintermsofbeta}) (up to order $O(p^{-1})$), and in the last line, we reinserted the expression for $\langle G^2 \rangle$ back into the right-hand side, and ignored $O(p^{-2})$ terms. Inserting these into Eq.~(\ref{avresintermsofbeta}), we obtain
\begin{align}
&\langle G \rangle \approx \frac{1}{\omega - \langle G \rangle} + \frac{\beta_4 \langle G\rangle^2}{p(\omega - \langle G \rangle)^3} + \frac{\beta_4^2 \langle G\rangle^2}{p^2(\omega - \langle G \rangle)^7}  \nonumber \\
&+  \frac{\beta_6\langle G\rangle^3}{p^2(\omega - \langle G \rangle)^4} + \frac{3 \beta_4^2 \langle G\rangle^4}{p^2(\omega - \langle G \rangle)^5}.\label{secondorder2}
\end{align}    
We have thus succeeded in finding an approximate expression for $\langle G\rangle$ in terms of only itself. We can simplify the expression somewhat by reinserting the right-hand side back into itself, ignoring terms that contribute to $O(p^{-3})$. To first order in $p^{-1}$, we obtain $\langle G \rangle \approx (\omega-\langle G \rangle)^{-1} + \beta_4 p^{-1} \langle G\rangle^2(\omega - \langle G \rangle)^{-3} \approx (\omega-\langle G \rangle)^{-1} + \beta_4 p^{-1} \langle G\rangle^5$, which then gives us and approximation for $(\omega-\langle G \rangle)^{-1}$. Inserting this back into Eq.~(\ref{secondorder2}), we finally arrive at Eq.~(\ref{firstmoment}).

\section{Laplace's method applied to Eq.~(\ref{fullsol})}\label{appendix:laplace}

We wish now to approximate Eq.~(\ref{fullsol}) for large and small $x$ using Laplace's method. We begin with the large-$x$ case. We define
\begin{align}
P_\Sigma(X,Y) = \exp\left[ F(X,Y) \right] ,
\end{align}
and we keep our reasoning as generally applicable to different forms of $F(X,Y)$ as possible. We focus on the two contributions to the integrand in Eq.~(\ref{fullsol}) individually. For the integrand involving $Y_+$, the maximum occurs at $dL/dX = 0$ where
\begin{align}
L(X, Y_+(X)) =& F(X, Y_+(X)) \nonumber \\
&+ \log\left[\frac{(Y_+^2 + (X-\omega)^2)^2}{Y_+^2 - (X-\omega)^2} \right]. \label{lfunct}
\end{align}
One obtains
\begin{align}
\frac{dL}{dX} =& \frac{\partial F}{\partial X} + \frac{\partial F}{\partial Y_+}\frac{\partial Y_+}{\partial X} + \frac{2 Y_+[Y_+^2 -3(X-\omega)^2]}{Y_+^4 - (X-\omega)^4}\frac{d Y^+}{dX} \nonumber \\
&+ \frac{2(X-\omega)[3Y_+^2- (X-\omega)^2]}{Y_+^4 - (X-\omega)^4},\label{dlbydx}
\end{align}
where we also have
\begin{align}
Y_+ &= \frac{1 + \sqrt{1 - 4\pi^2 x^2 (\omega-X)^2}}{2\pi x}, \nonumber \\
\frac{d Y_+}{d X} &=   \frac{2 \pi x(\omega- X)}{\sqrt{1 - 4 \pi^2x^2 (\omega-X)^2}} .\label{derivatives1}
\end{align}
We note that the integration range in Eq.~(\ref{fullsol}) ensures the reality of $Y_+(X)$. So, necessarily, $(X-\omega)^2<1/(4x^2)$ and therefore $Y_+ \sim 1/x$ for large $x$. This means that we have 
\begin{align}
\frac{dL}{dX}\approx&  \frac{\partial F}{\partial X} + a\frac{\partial F}{\partial Y_+}x (X-\omega) \nonumber \\
&+ b(X-\omega) x^2 + c x^4(X-\omega)^3,
\end{align}
where $a$, $b$ and $c$ are constants of order $O(x^0)$ and $O((X-\omega)^0)$. Let us now attempt to find the order of magnitude of $(X-\omega)$ in terms of $x$ by using $dL/dX = 0$. Assuming that $\partial F/\partial X \sim \partial F/\partial Y_+ \sim x^0$, we see that we must have $(X-\omega) \sim1/x^2$ in order for the solution of $dL/dX = 0$ to be consistent. This means that the maximum of the integrand for large $x$ is given by
\begin{align}
Y_+(X^\star) = \frac{1}{\pi x} + O(x^{-3}), \hspace{1cm} X^\star = \omega + O(x^{-2}). \label{saddlepoint}
\end{align}
One can verify the assumption that indeed $\partial F/\partial X \sim \partial F/\partial Y_+ \sim x^0$ at the maximum \textit{a posteriori} for the approximation for $P_\Sigma(X,Y)$ discussed in Section \ref{section:selfenergydist}. By similar reasoning, we find that the contribution to the integrand in Eq.~(\ref{fullsol}) involving $Y_-$ must be subleading for large $x$, since one has instead $Y_- \sim x^{-3}$ with $(X-\omega) \sim1/x^2$. 

Now that we understand where the integrand of Eq.~(\ref{fullsol}) is maximum, we can perform Laplace's approximation. That is, we expand the logarithm of the integrand in Eq.~(\ref{fullsol}) around the maximum up to second order. This provides a Gaussian approximation for the integral, which we then evaluate. We have 
\begin{align}
\frac{d^2 F(X, Y_+(X))}{dX^2}  =& \frac{\partial^2 F}{\partial X^2} + 2\frac{\partial^2F}{\partial Y \partial X} \frac{d Y_+}{dX} + \frac{\partial^2 F}{\partial Y^2} \left( \frac{d Y_+}{d X} \right)^2 \nonumber \\
&+ \frac{\partial F}{\partial Y}\frac{d^2 Y_+}{d X^2}, \label{secondderivativef}
\end{align}
where in addition to Eq.~(\ref{derivatives1}) we have
\begin{align}
\frac{d^2 Y_+}{d X^2} &=  \frac{2\pi x}{\left(1 - 4x^2 \pi^2 (\omega-X)^2\right)^{3/2}} . 
\end{align}
Assuming that the power series for $F$ has no terms proportional to $Y^{-r}$ for $r>0$, we see that the largest power of $x$ that does not vanish when we set $X = \omega$ in Eq.~(\ref{secondderivativef}) comes from the term proportional to $ \frac{d^2 Y_+}{d X^2} $. This is proportional to $x$. 

However, if we examine the second derivative of the other term in Eq.~(\ref{lfunct}), we obtain at $X = \omega$
\begin{align}
\frac{d^2}{dX^2}\log\left[\frac{(Y_+^2 + (X-\omega)^2)^2}{Y_+^2 - (X-\omega)^2} \right] \bigg\vert_{X = \omega} = 2 \pi^2 x^2.
\end{align}
This therefore provides the dominant term in $L''$ for large $x$, which is proportional to $x^2$, and this ought to be the case generically for any $F(X,Y)$ that does not have poles at $Y = 0$. Simple application of Laplace's method then yields Eq.~(\ref{asymp}). 

One notes that this reasoning breaks down if the maximum in Eq.~(\ref{saddlepoint}) is somehow inaccessible. This is the case, for example, when $\beta_4/p = O(1/N)$ 
or when $X$ and $Y$ are otherwise linearly dependent, as was discussed in Section \ref{section:breakdown}. 

It remains for us to understand the behaviour for small $x$. Again, since $P_\mathrm{LDoS}(x)$ is small for small $x$ in comparison to the peak (i.e. small values of $x$ are `rare events'), we expect the integrand to be dominated by contributions near the maximum value, and so Laplace's method again ought to yield the asymptotic behaviour. We focus first on the second contribution to the integrand in Eq.~(\ref{fullsol}), which involves $Y_-$.

From Eq.~(\ref{dlbydx}), and using that for small $x$ we have $Y_- \approx \pi x(\omega-X)^2$ and $dY_-/dX \approx 2 \pi x (X-\omega)$ (assuming $\omega - X \sim x^0$), we have that 
\begin{align}
\frac{dL}{dX} \approx \frac{\partial F}{\partial X} + \frac{2}{X-\omega} .
\end{align}
Setting $\frac{dL}{dX} = 0$ in this expression yields the maximum $X^\star$, at which indeed $\omega - X \sim x^0$. We may now evaluate the second derivative at this value, obtaining for small $x$
\begin{align}
\frac{d^2L}{dx^2} \approx  \frac{\partial^2 F}{\partial X^2} + \frac{2}{(\omega-X)^2} \sim x^0.
\end{align}
If we instead consider the contribution involving $Y_+$, one finds $Y_+ \sim 1/x$ but $dY_+/dX \sim x(\omega - X)$. We thus have $X-\omega \sim 1/x$. In this case, the contribution to $P_\mathrm{LDoS}(x)$ is then dominated by the factor $P_\Sigma(X, Y_+)$, which is exponentially small. The dominant contribution thus comes from the term involving $Y_-$. We therefore finally obtain via Laplace's method for $x \to 0$
\begin{align}\label{smallxapprox}
P_\mathrm{LDoS}(x) \approx (\omega-X^\star)^2  \sqrt{\frac{2\pi^3}{L''(X^\star)}}P_\Sigma[X^\star, \pi x (\omega-X^\star)^2] ,
\end{align}
which is constant in $x$.

\section{Eigenvector support set}\label{appendix:support_set}
Consider the set of eigenvector components for a state $\nu$, $\psi^{(\nu)}_i$. We order them in terms of magnitude such that
\begin{align}
\vert \psi_1^{(\nu)} \vert^2 \geq \vert \psi_2^{(\nu)} \vert^2 \geq \vert \psi_3^{(\nu)} \vert^2 \geq \cdots \geq \vert \psi_N^{(\nu)} \vert^2.
\end{align}
The support set \cite{kravtsov2018nonergodic} is defined as the least number $M_\epsilon$ of these components that are necessary to satisfy the normalisation constraint up to some prescribed accuracy $\epsilon$. That is,
\begin{align}
\sum_{i = 1}^{M_\epsilon} \vert \psi_i^{(\nu)} \vert^2 \leq 1-\epsilon<\sum_{i = 1}^{M_\epsilon+1} \vert \psi_i^{(\nu)} \vert^2. \label{supportsetdef}
\end{align} 
Taking the limit $N \to \infty$ in Eq.~(\ref{supportsetdef}), we have
\begin{align}
N\int_{y_\epsilon}^N dy \, P_\psi(y) &= M_\epsilon, \nonumber \\
\int_{y_\epsilon}^N dy \,y P_\psi(y) &= 1-\epsilon.
\end{align}
Let us compute $M_\epsilon$ for the eigenvectors of the random matrix ensembles that we have studied. As long as $y_\epsilon$ is in the power-law tail of the eigenvector distribution (as it should be for small $\epsilon$), we find 
\begin{align}
\frac{B}{2y_\epsilon^2}= \frac{M_\epsilon}{N}, \hspace{1cm} \frac{B}{y_\epsilon} = 1-\epsilon.
\end{align}
Eliminating $y_\epsilon$ we therefore find for the support set
\begin{align}
M_\epsilon = \frac{N(1-\epsilon)^2}{2 B}. \label{nongausssupportset}
\end{align}
Crucially, we must understand the scaling of $B$ with $N$ to obtain the scaling of the support set with $N$. We can obtain an estimate the constant $B$ as follows. We assume that the distribution $P_\psi(y)$ follows a scaled Porter-Thomas form up to a value $y = y_c$, beyond which we have the power law behaviour $P_\psi(y) = B/y^3$. By requiring continuity and normalisation of both the distribution $P_\psi(y)$ and the eigenvectors, we arrive at the conclusion that $B\sim N^0$. Therefore, we find that the support set $M_\epsilon \sim N$ is extensive. We note that in reality the constant $B$ varies between weight distributions and is dependent on other system parameters such as $p$ and $\omega$.

\section{Behaviour close to the spectral edge and beyond}\label{appendix:edge}

Following Ref. \cite{kravtsov2018nonergodic}, let us suppose that the imaginary part of the self-energy is small in comparison to the real part. This is obviously the case at the edge of the spectrum where the eigenvalue density approaches zero. We therefore approximate [c.f. Eq.~(\ref{imgir})]
\begin{align}
\mathrm{Im} G_{ii} \approx \frac{Y_i}{(\omega-X_i)^2} .
\end{align}
Let us briefly examine the relationship between $Y_i$ and $X_i$. From Eqs.~(\ref{genfunctsigma}), we see that, to leading order in $p^{-1}$, $X$ and $Y$ are linearly dependent since the fluctuations of $G_{ii}$ go as $p^{-1}$ (this is also discussed in Section \ref{section:breakdown}). For this reason, we may write $Y = z \langle \mathrm{Im} G \rangle$ and $X= z \langle \mathrm{Re} G \rangle$, where $z$ is distributed according to
\begin{align}
P_z(z) = \frac{1}{\sqrt{2 \pi \beta_4/p}}e^{-\frac{(z-1)^2}{2 \beta_4/p}}.
\end{align}
We note that this approximate linear dependence is not used in the previous calculation of $P_\mathrm{LDoS}(x)$ in Sections \ref{section:selfenergydist} and \ref{section:ldosdist}. In fact, the slight fluctuations about this solution are crucial to obtain large values of $x$ when $\langle Y\rangle$ is not small, as was discussed in Section \ref{section:breakdown}.  

With this recasting of $X$ and $Y$ in terms of the single variable $z$, and writing $\langle Y \rangle = \pi \rho$, we obtain
\begin{widetext}

\begin{align}
P(\mathrm{Im}G_{ii} = x) &\approx \int dz\,\delta\left(x - \frac{\pi \rho z}{(\omega - \langle X \rangle z )^2}\right) \frac{1}{\sqrt{2 \pi \beta_4/p}}e^{-\frac{(z-1)^2}{2 \beta_4/p}} \nonumber \\
&= \left\vert \frac{\sqrt{\pi \rho(\pi \rho + 4 \langle X \rangle \omega x)}- \pi \rho}{2 \langle X\rangle^2 x^2 } - \frac{\omega \sqrt{\pi \rho}}{\langle X \rangle x \sqrt{\pi \rho + 4 \langle X \rangle \omega x}}\right\vert \frac{\exp\left[-\frac{(z(x)-1)^2}{2\beta_4/p}\right]}{ \sqrt{2 \pi \beta_4/p}}, \nonumber \\
z(x) &= \frac{\pi \rho + 2 \langle X \rangle \omega x - \sqrt{\pi \rho (\pi \rho + 4 \langle X \rangle \omega x)}}{2 \langle X \rangle^2 x} .
\end{align}
\end{widetext}
In the limit $x\to \infty$, we have
\begin{align}
P(\mathrm{Im}G_{ii} = x) &\approx  \frac{\sqrt{\pi \rho \omega}}{2(\langle X \rangle x)^{3/2}} \frac{\exp\left[-\frac{(\omega/\langle X\rangle-1)^2}{2\beta_4/p}\right]}{ \sqrt{2 \pi \beta_4/p}} .
\end{align}
In reality however, this behaviour will only be exhibited asymptotically when we are exactly at the spectral edge. 

We now show that this same power-law behaviour is preserved in the point-like part of the spectrum (i.e. where the eigenvectors are localised). In this region of the spectrum, the resolvent has simple poles at various locations $E_\nu$, corresponding to the outlier eigenvalues generated by the hubs of the network. One can indeed show that these poles correspond to localised states \cite{valigi2025eigenvalue}, and we can also characterise the LDoS distribution in this region of the spectrum following Ref. \cite{biroli2018delocalization}. For small $\epsilon$ we can write the following using the pole-like nature of the local resolvent above (when $\omega \approx E_\nu$)
\begin{align}
\mathrm{Im}G_{ii}(\omega) = \frac{\vert \psi_i^{(\nu)}\vert^2\epsilon}{(\omega  - E_\nu)^2 + \epsilon^2}. 
\end{align}
By a simple change of variables, one thus finds
\begin{align}
P_\mathrm{LDoS}(x) \propto \frac{\sqrt{\epsilon}}{x^{3/2}}
\end{align}
until a cut-off of $x \sim \frac{1}{\epsilon}$.

\end{document}